# Statistical robustness analysis of fractional and integer order PID controllers for the control of a nonlinear system

J. Viola and L. Angel

*Abstract*— This paper presents the design and robustness analysis of fractional and integer order PID controllers for the control of a non-linear industrial process in the presence of parametric uncertainness and external disturbances. The nonlinear system is linearized using an input-output linearization technique. Three controllers were designed for the linearized system, an integer order PID controller, a fractional PID controller, and a SIMC PID controller. The robustness analysis of the proposed controllers is based on a $2^3$ factorial experimental design. The input factors for the experiment are the uncertainty in gains of the plant, the presence of random noise in the feedback loop, and the existence of external perturbations. The outputs of the experiment measure the performance of each controller through the time step response and the control action of each controller using the mean value and the standard deviation. The obtained results show that the fractional order PID controller has better performance in the presence of the analyzed experimental factors, especially in the control action indicating greater robustness and lower energy consumption.

## I. INTRODUCTION

PID controllers are widely employed in the industry for process control due to its simplicity for tuning, design, and implementation. One example is the Skogestad Internal Model control (SIMC) PID controller, which is an excellent alternative to control industrial processes due to the simplicity of tuning, and its good external disturbance rejection [1]-[3].

Since the last few years, a better understanding of fractional order calculus has extended the use of fractional order operators to the modeling and control of industrial processes with better results against integer-order models [4], [5]. Likewise, the use of fractional operators in the system modeling and representation makes possible obtaining better modeling of the dynamical behavior of a system without using high order approximations [4]-[7]. In addition, fractional order operators applied in controller design increases the number of tuning parameters of the controller, making easier to obtain the desired response of a system, and improve the controller robustness in the presence of external disturbances, parametric uncertainness, and random noises [8]-[20]. On the other hand, the presence of external perturbations or parametric uncertainties in the process model has a significant effect on the control system behavior affecting the desired performance of the system. So that, advanced control strategies should be employed to deal with this undesired behavior, which has a greater complexity of the controller tuning and practical implementation. Indeed, quantifying the effect of these external disturbances and parametric uncertainness will contribute to understand better the system dynamic and evaluate the robustness of the employed control strategies to these undesired behaviors.

This paper presents the statistical robustness analysis of a fractional order PID controller (FOPID), an integer order PID controller (IOPID), and a SIMC PID controller applied to the control of a non-linear level system with parametric uncertainty and external disturbances. This analysis employs a $2^3$ factorial experimental design methodology, which factors are the presence or absence of parametric uncertain in the plant gains, the presence or absence of random noise in the feedback loop, and the presence or absence of external disturbances in the control action of each controller. The factorial experimental design outputs are the integral square error (*ISE*) and the standard deviation for the step response of the system. Likewise, the control action is evaluated through its average value and its standard deviation. Then an ANOVA analysis is performed to determinate the influence of each factor for each experimental condition and each controller.

The main contribution of this paper is the design of a FOPID controller for a non-linear system, and the use of a factorial experimental design $2^3$ to perform a statistical robustness analysis of the FOPID, IOPID and SIMC PID controllers in the presence of external disturbances and parametric uncertainness.

This paper is structured as follows. First, a review of the basic concepts of fractional calculus is presented. Second, the design technique for FOPID, IOPID, and SIMC PID controllers is developed. Third, the non-linear system model and its input-output linearization are presented. Fourth, the controllers design is presented. Fourth, the factorial experimental design $2^3$ is developed to analyze the robustness of each controller. Finally, conclusions are presented.

## II. PRELIMINARY CONCEPTS

*A. Fractional operators*

In the theory of fractional calculus [4], [5], the notation used to represent the integral and the fractional derivative is

J. Viola is with the Universidad Pontificia Bolivariana, Seccional Bucaramanga, Colombia, (email: jairoviola92@hotmail.com).

L. Angel is with the Universidad Pontificia Bolivariana, Seccional Bucaramanga, Colombia, (email: luis.angel@upb.edu.co).

$$D_t^{\pm\alpha}f(t) = \begin{cases} \dfrac{d^\alpha}{dt^\alpha}f(t), & \alpha < 0 \\ f(t), & \alpha = 0 \\ \dfrac{d^{-\alpha}}{dt^{-\alpha}}g(t) = I^{-\alpha}f(t), & \alpha \geq 0 \end{cases} \quad (1)$$

where $\alpha$ and $t$ are the lower and upper limit for the operation. $\alpha$ and $-\alpha$ are the non-integer order of the derivative and the fractional integral respectively, and $f(t)$ is the function to be integrated or differentiated. Based on the general fractional order operators (1), the derivative and integral could be defined according to the Riemann–Liouville, and the Grünwald-Letnikov definitions given by (2) and (3), where $t = hk$, k is the number of steps, and h the step size.

$$D^{-n}f(x) = \frac{1}{\Gamma(\alpha)}\int_a^x f(t)(x-t)^{\alpha-1}dt \quad (2)$$

$$f^\alpha(t) = \lim_{h\to 0}\frac{1}{h^\alpha}\sum_{j=0}^{\infty}(-1)^j \binom{\alpha}{r} f(t-jh) \quad (3)$$

With zero initial conditions, the Laplace Transform for the non-integer operators (2), (3) is defined as

$$L\{D^{\pm\alpha}f(t)\} = s^{\pm\alpha}f(s) \quad (4)$$

According to (2) and (3), the FOPID controller integro-differential equation is defined by (5)

$$u(t) = K\left(e(t) + \frac{1}{T_i}D^{-\lambda}e(t) + T_d D^\mu e(t)\right) \quad (5)$$

where $K$ is the proportionality constant; $\tau_i$ is the integral time constant; $\tau_d$ is the derivative time constant, $\lambda$ refers to the non-integer order of the integrator, and $\mu$ refers to the non-integer order of the derivative. Based on (5) and applying the Laplace transform described in (4), the transfer function of the FOPID controller is denoted by (6).

$$G_c(s) = \frac{U(s)}{E(s)} = K\left(1 + \frac{1}{T_i}s^{-\lambda} + T_d s^\mu\right) \quad (6)$$

### III. CONTROLLERS DESIGN TECHNIQUES

#### A. FOPID controller

The design of fractional controllers employing frequency domain techniques looks for providing robustness to the controller in the presence of parametric uncertainties of the process and the presence of external perturbations [13]. The methodology for the frequency domain design of fractional order PID controllers should satisfy the following six conditions.

- Phase margin (pm):

$$\arctan\left(G_c(jw)G_p(jw)\right) = -\pi + pm \quad (7)$$

- Gain crossover frequency (Wc):

$$|G_c(jw)G_p(jw)| = 0\ DB \quad (8)$$

- Robustness against variations in plant gains:

$$\frac{d}{dw}arctan(G_c(jw)G_p(jw)) = 0 \quad (9)$$

- Rejection of high frequency noise:

$$\left|\frac{G_c(jw)G_p(jw)}{1 + G_c(jw)G_p(jw)}\right| = B\ dB \quad (10)$$

- Rejection of output perturbations:

$$\left|\frac{1}{1 + G_c(jw)G_p(jw)}\right| = A\ dB \quad (11)$$

- Controller saturation: This condition is important, since the actuators in real systems have physical limits, which cannot be exceeded. Accordingly, the control action will be limited between 0V-10V considering the industry standards.

Considering that equations set (7)-(11) is nonlinear, am optimization algorithm should be employed to find the optimal values for the FOPID controller terms. In this case, (8) is considered as the cost function of the system and (7), (9)-(11) are considered as the restrictions of the optimization problem. In addition, FMINCON function of Matlab is employed to find the FOPID controller terms based on the frequency domain specs as the phase margin, the gain margin, and the gain crossover frequency.

#### B. SIMC PID design technique

The SIMC tuning technique is based on the theory of internal model control (IMC) and proposes a simple methodology that allows the tuning of PI and PID industrial controllers [14]. This methodology has certain advantages for the controller design, as the simplicity of the parameters calculation of the PI and PID controllers for first and second-order systems ensuring the robustness of the system against setpoint changes, the presence of external perturbations and random noise. For a second-order system with dead time given by (12), the parameter of a SIMC PID can be calculated employing by (13).

$$Gp(s) = \frac{k}{(\tau_1 s + 1)(\tau_2 s + 1)}e^{-\theta s} \quad (12)$$

$$K_p = \frac{1}{k} * \frac{1}{\tau_c + \theta}$$

$$\tau_i = \min(\tau_1, 4(\tau_c + \theta)) \quad (13)$$

$$\tau_d = \tau_2$$

Notice that (13), has only one tuning parameter $\tau_c$. [14] suggests that this value should be equal to $\theta$ to ensure a robust control with good tracking of the reference changes.

### C. Design of IOPID controllers based on the frequency domain

The value of the parameters of the IOPID controller are found employing the same optimization methodology proposed for the FOPID controller in Section IIIA.

## IV. SYSTEM TO BE CONTROLLED AND DESIGN OF CONTROLLERS

For the robustness study of the FOPID, SIMC PID and IOPID controllers, we have used a second-order non-linear system which is linearized using input-output linearization technique. The design of the FOPID controller uses the solution of (7) – (11) and the ISE (integral of squared error) criterion as an optimization criterion. The SIMC PID controller uses the solution of (13). Finally, the design of the IOPID controller uses the solution of (7)-(9) and the ISE criterion as an optimization criterion.

### A. Second-order non-linear system

The plant corresponds to a system of two non-interacting tanks for which the state model is given by

$$\begin{bmatrix} \dot{x}_1(t) \\ \dot{x}_2(t) \end{bmatrix} = \begin{bmatrix} -\sqrt{x_1(t)} \\ \sqrt{x_1(t)} - \sqrt{x_2(t)} \end{bmatrix} + \begin{bmatrix} 1 \\ 0 \end{bmatrix} u(t) \quad (14)$$

where $x_1(t)$ and $x_2(t)$ refer to the levels in tanks 1 and 2, respectively, and u(t) refers to the input flow. In general form, (14) can be expressed as

$$\dot{x} = f(x) + g(x)u \quad (15)$$

where:

$$f(x) = \begin{bmatrix} -\sqrt{x_1(t)} \\ \sqrt{x_1(t)} - \sqrt{x_2(t)} \end{bmatrix} \text{ and } g(x) = \begin{bmatrix} 1 \\ 0 \end{bmatrix} \quad (16)$$

The output of the system is defined as follows:

$$y(t) = x_2(t) = h(x). \quad (17)$$

The control system develop starts with the input-output linearization using Lie algebra [21], [22]. This technique proposes the transformation of the non-linear system into a linear system and a control system, followed by a change in the coordinate system. The law of linearization control is given by

$$u = \frac{1}{l_g l_f^{\varphi-1} h(x)} (-l_f^\varphi h(x) + v) \quad (18)$$

The resulting linear model of order, with $v$ as input and y(t) as output, is

$$\frac{d^\varphi y(t)}{dt^\varphi} = v(t) \quad (19)$$

where $\varphi$ is denominated the relative degree of the system. If (19) has a stable dynamic of zero and the relative degree of the system is $\varphi$, the law of state feedback is

$$u = \frac{1}{l_g l_f^{\varphi-1} h(x)} \left( -\sum_{j=0}^{\varphi} \beta_j l_f^j h(x) + v \right) \quad (20)$$

The control law (20) stabilizes the system exponentially with the following characteristic polynomial:

$$\beta_\varphi s^\varphi + \beta_{\varphi-1} s^{\varphi-1} + \cdots + \beta_1 s + \beta_0 \quad (21)$$

For (14), there are the following Lie derivatives:

$$\begin{aligned} l_f h(x) &= \frac{\partial h(x)}{\partial x} f(x) = \sqrt{x_1(t)} - \sqrt{x_2(t)} \\ l_g h(x) &= \frac{\partial h(x)}{\partial x} g(x) = 0 \\ l_f^2 h(x) &= \frac{\partial l_f h(x)}{\partial x} f(x) = -\frac{1}{2} \sqrt{\frac{x_1(t)}{x_2(t)}} \\ l_g l_f h(x) &= \frac{\partial l_f h(x)}{\partial x} g(x) = \frac{1}{2\sqrt{x_1(t)}} \end{aligned} \quad (22)$$

Given that $l_g l_f h(x) \neq 0$, the relative order of the non-linear system is $\varphi = 2$. From the answer in the open loop against the unit step input for the non-linear system, we have that the settling time is 14 s without overshoot. to maintain the same transitory characteristics, the desired characteristic polynomial for the linearized system is

$$\begin{aligned} s^2 + \beta_1 s + \beta_0 &= s^2 + 1.66s + 0.666 \\ &= (s + 0.8471)(s + 0.7864) \end{aligned} \quad (23)$$

Using (23), the state feedback law (20) would be given by

$$u = \frac{1}{l_g l_f h(x)} [-\beta_0 h(x) - \beta_1 l_f h(x) - l_f^2 h(x) + v] \quad (24)$$

Using (22) and (23) in (24), we obtain the following:

$$\begin{aligned} u = 2\sqrt{x_1(t)} [&-0.666 x_2(t) - \\ &1.66[\sqrt{x_1(t)} - \sqrt{x_2(t)}] + \frac{1}{2}\sqrt{\frac{x_1(t)}{x_2(t)}} + v] \end{aligned} \quad (25)$$

Substituting (25) into (15), we obtain the following linearized system:

$$G_p(s) = \frac{2}{s^2 + 1.66s + 0.666} e^{-0.5s} \quad (26)$$

The linearized system (26) is employed to design the FOPID, IOPID and SIMC PID controllers. The desired specifications for the FOPID and IOPID controllers are a gain margin of $10\ dB$, a phase margin of $75°$, and a gain crossover frequency of $1.94\ rad/s$. Table I presents the obtained values for the FOPID, IOPID, and SIMC PID controllers as well as the ISE performance index for the step response of the system, and the average value of the control action.

TABLE I. FOPID, IOPID, AND SIMC PID CONTROLLER TERMS FOR NON-LINEAR SYSTEM

| Controller | $K_p$ | $\tau_i$ | $\tau_d$ | $\lambda$ | $\mu$ | ISE | Average control action |
|---|---|---|---|---|---|---|---|
| FOPID | 0.46 | 0.64 | 3.2 | 0.85 | 0.67 | 0,73 | 0,29 |
| SIMC PID | 4.94 | 10.2 | 0.002 | 1 | 1 | 1,56 | 0,76 |
| IOPID | 0.76 | 1.4 | 0.003 | 1 | 1 | 0,82 | 0,33 |

As can be observed, the step response using the FOPID controller exhibits better behavior with a lower *ISE* value than the SIMC PID and IOPID controllers. In addition, the FOPID controller has a lower average value for the control action compared with the IOPID and SIMC PID controllers. Figure. 1 shows the step response, and the control action of the system. As can be observed, the maximum overshoot is reached by the SIMC PID controller, and the settling time is similar for all controllers. The control action does not reach the saturation level when the FOPID controller is used. while IOPID and SIMC PID reach saturation values in the control action.

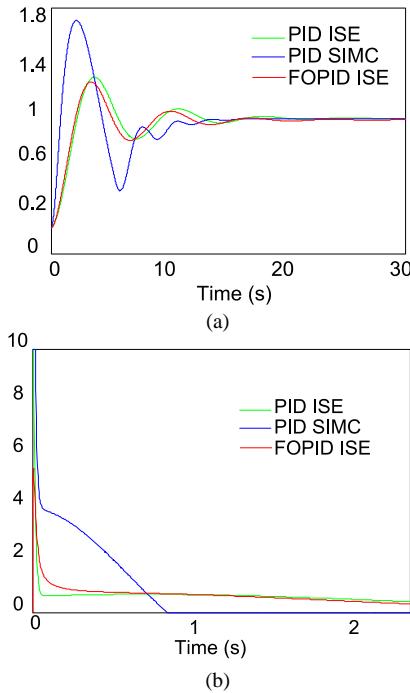

Figure 1. Non-linear system a) step response, b) control action employing the FOPID, IOPID, and SIMC PID controllers

## V. ROBUSTNESS ANALYSIS BASED ON $2^3$ EXPERIMENTAL DESIGN

The robustness of a control system is defined as the capacity of the controller to ensure robust stability and robust performance specifications improving the rejection of external perturbations and minimizing the random noise [14]. This paper proposes the robustness evaluation of the controllers presented in Section IV, considering their dynamic behavior in the presence of parametric uncertain in the plant gains, random noise in the feedback loop, and external disturbances in the control action. For this reason, a factorial $2^3$ experimental design is proposed, in which we consider the presence or absence of the following factors: A: uncertainty in the gains of the plant of +100%, B: random noise of $\pm 10\%$ and C: external perturbation of +20%. The description and levels of these factors are shown in Table 2.

The robustness will be evaluated through the dynamic behavior of the step response of the system and the control action. The outputs of the factorial experimental design are the *ISE* criterion, and the standard deviation for the input step response of the system. For the control action, the factorial experimental outputs are the mean value, and the standard deviation. Two replications of the experiment are performed to obtain a good data representation from a statistical point of view. Table III to Table V show the effect of each factor and their combinations on the outputs mentioned previously for the system (14) employing the FOPID, IOPID, and SIMC PID controllers. As can be observed, the highest *ISE* value is found when the SIMC PID controller is employed (4.639), and the lowest value is obtained for the FOPID controller (0.49). The lowest standard deviation is presented by the FOPID controller (0.13), and the greatest standard is presented when the SIMC PID controller is employed (0.393). Regarding to the control action, the maximum mean value is reached by SIMC PID controller (1.81), and the minimum mean value (0.11) for the FOPID controller, which is significantly lower compared with the SIMC PID controller. The control action has the lowest standard deviation value when FOPID controller is used (0.21), and its maximum value is seen when using the SIMC PID and IOPID controllers (1.173). Figure 2 shows the combined effect of the 3 factors on the input step response and the control action for each of the controllers. The perturbation has been applied to control action of the system at $t = 15s$. As can be observed, the FOPID controller has the best performance in the step response, and better control action demonstrated with a lower ISE index and a lower mean value of the control action.

TABLE II. $2^3$ FACTORIAL EXPERIMENTAL DESIGN FACTORS

| Level | A: Uncertainty in gain of the plant | B: random noise | C: external disturbance |
|---|---|---|---|
| 0 | 100% | 0 | 0 |
| 1 | 200% | $\pm 10\%$ | +20% |

TABLE III. EFFECT OF FACTORS ON THE LINEARIZED SECOND-ORDER SYSTEM WITH THE FOPID CONTROLLER

| FACTOR | | | STEP RESPONSE | | CONTROL ACTION | |
|---|---|---|---|---|---|---|
| C | B | A | ISE | SD | MEAN VALUE | SD |
| 0 | 0 | 0 | 0,49 | 0,13 | 0,11 | 0,208 |
| 0 | 0 | 1 | 0,49 | 0,13 | 0,11 | 0,208 |
| 0 | 1 | 0 | 0,73 | 0,15 | 0,292 | 0,21 |
| 0 | 1 | 1 | 0,81 | 0,163 | 0,323 | 0,29 |
| 1 | 0 | 0 | 0,77 | 0,161 | 0,252 | 0,219 |
| 1 | 0 | 1 | 0,76 | 0,158 | 0,366 | 0,293 |
| 1 | 1 | 0 | 0,81 | 0,163 | 0,323 | 0,294 |
| 1 | 1 | 1 | 0,59 | 0,141 | 0,169 | 0,263 |

TABLE IV. EFFECT OF FACTORS ON THE LINEARIZED SECOND-ORDER SYSTEM WTH THE IOPID CONTROLLER

| FACTOR | | | STEP RESPONSE | | CONTROL ACTION | |
|---|---|---|---|---|---|---|
| C | B | A | ISE | SD | MEAN VALUE | SD |
| 0 | 0 | 0 | 0,867 | 0,172 | 0,296 | 0,438 |
| 0 | 0 | 1 | 0,866 | 0,170 | 0,720 | 0,743 |
| 0 | 1 | 0 | 0,536 | 0,137 | 0,155 | 0,437 |
| 0 | 1 | 1 | 0,826 | 0,168 | 0,337 | 0,433 |
| 1 | 0 | 0 | 0,867 | 0,172 | 0,296 | 0,438 |
| 1 | 0 | 1 | 0,913 | 0,175 | 0,667 | 0,739 |
| 1 | 1 | 0 | 0,913 | 0,175 | 0,667 | 0,739 |
| 1 | 1 | 1 | 0,826 | 0,168 | 0,337 | 0,433 |

TABLE V. EFFECT OF FACTORS ON THE LINEARIZED SECOND-ORDER SYSTEM AND THE SIMC CONTROLLER

| FACTOR | | | STEP RESPONSE | | CONTROL ACTION | |
|---|---|---|---|---|---|---|
| C | B | A | ISE | SD | MEAN VALUE | SD |
| 0 | 0 | 0 | 3,67 | 0,352 | 0,403 | 0,555 |
| 0 | 0 | 1 | 1,562 | 0,230 | 0,762 | 0,620 |
| 0 | 1 | 0 | 1,597 | 0,232 | 0,733 | 0,629 |
| 0 | 1 | 1 | 1,976 | 0,257 | 1,699 | 1,148 |
| 1 | 0 | 0 | 1,976 | 0,257 | 1,699 | 1,148 |
| 1 | 0 | 1 | 4,639 | 0,393 | 1,136 | 1,041 |
| 1 | 1 | 0 | 1,976 | 0,257 | 1,699 | 1,148 |
| 1 | 1 | 1 | 1,944 | 0,256 | 1,816 | 1,173 |

## VI. RESULTS AND DISCUSSION

The factorial experimental design $2^3$ is applied to each one of experiment outputs, two of them evaluate the robustness of the system through the input step response (ISE and standard deviation), and the remain outputs are employed to evaluate the robustness of the system through the control action (mean value and standard deviation). The ANOVA analysis allows finding the percentage of influence of each factor (perturbations and uncertainties) over each one of the analyzed outputs, and for the FOPID, IOPID, and SIMC PID controllers, which is performed using the experimental design software Design Expert 7.0. Considering that the experiment has four outputs per controller and three controllers are evaluated, the total number of experiments performed is 24. The results obtained for each controller are shown in Table 6 which shows the influence percentage of each factor and their combined effects for the input step response of the system (ISE and standard deviation) and the influence percentage of each factor, and their combined effects for the control action (mean value and standard deviation). The measurement factor (MF) is defined to study the robustness of each controller against the presence of parametric uncertainty in the gains of the plant, random noise in the feedback loop, the presence of external perturbations, and its combined effects. The measurement factor is given by (27) where *experimental data* refers to the data shown in Table III to Table V for each one of the factors, and their combined effects, and *influence %* is the result of the factorial design shown in Table 6.

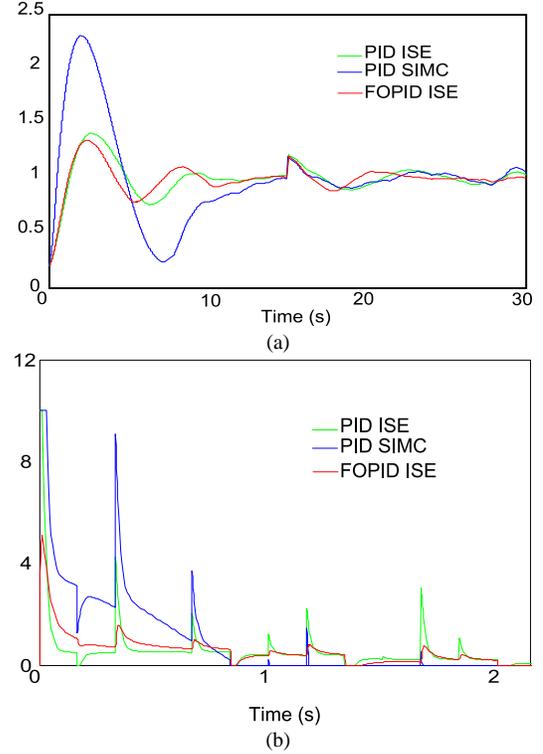

Figure 2. Combined response of the 3 factors on the non-linear second-order system for each control system: a) input step response b) control action.

$$MF = Experimental\ data * \frac{influence\ \%}{100\%} \qquad (27)$$

The MF values are shown in Fig 3 and Fig 4 for the step response and control action. As seen in Fig 3(a) and Fig 3(b), the *ISE* and standard deviation of the step response of the system using the FOPID and the IOPID controllers is affected by uncertainty in the gains of the system and random noise. Based on Fig 4, The mean value and standard deviation of the control action using the FOPID controller are not greater affected by any of the factors (variation in plant gains, random noise in feedback loop and presence of external perturbation) or their possible combinations, while the SIMC PID and IOPID controllers are more strongly affected by the noise factor (B). According to this analysis, the FOPID controller exhibits the best performance for the step response of the system, and the controller action, because *ISE* and standard deviation are smaller than in IOPID and SIMC PID controllers. So that, the FOPID controller is more robust in the presence of external disturbances and parametric uncertainness than the IOPID and the SIMC PID controllers.

TABLE VI. RESULTS OF FACTORIAL ANALYSIS: EFFECT OF EACH FACTOR ON STEP RESPONSE AND CONTROL ACTION FOR EACH CONTROLLER (%)

| FACTOR | Step response ISE (%) | | | Step response Standard deviation (%) | | | Average controller action (%) | | | Controller action Standard deviation (%) | | |
|---|---|---|---|---|---|---|---|---|---|---|---|---|
| | FOPID | PID SIMC | IOPID | FOPID | PID SIMC | IOPID | FOPID | PID SIMC | IOPID | FOPID | PID SIMC | IOPID |
| A | 45,761 | 60,561 | 39,534 | 46,368 | 62,801 | 40,994 | 47,567 | 9,505 | 10,329 | 1,592 | 1,562 | 0,011 |
| B | 22,089 | 0,006 | 32,512 | 18,641 | 0,000 | 29,110 | 30,478 | 83,708 | 84,623 | 93,456 | 97,416 | 99,904 |
| AB | 12,026 | 8,513 | 7,134 | 14,654 | 7,547 | 7,487 | 0,660 | 0,865 | 0,177 | 2,084 | 0,182 | 0,013 |
| C | 10,141 | 2,730 | 17,213 | 11,215 | 4,629 | 19,482 | 2,732 | 0,000 | 0,341 | 0,485 | 0,001 | 0,030 |
| AC | 3,099 | 9,419 | 1,173 | 2,845 | 8,523 | 1,103 | 9,099 | 2,529 | 2,238 | 0,626 | 0,332 | 0,012 |
| BC | 3,572 | 9,401 | 1,320 | 3,130 | 8,356 | 0,942 | 4,501 | 1,183 | 1,007 | 0,396 | 0,103 | 0,000 |
| ABC | 3,312 | 9,370 | 1,115 | 3,147 | 8,143 | 0,881 | 4,962 | 2,209 | 1,286 | 1,362 | 0,404 | 0,029 |

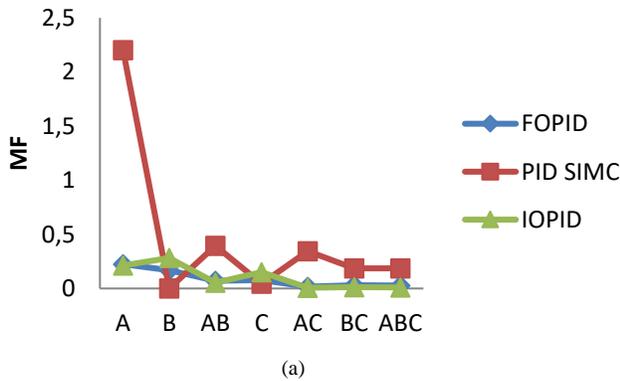
(a)

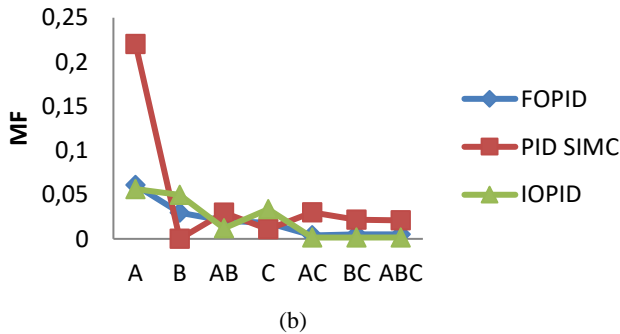
(b)

Figure 3. MF for robustness analysis of the input step response of the system: b) ISE and c) standard deviation.

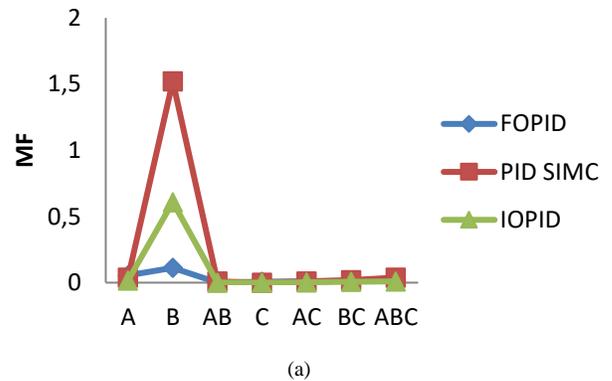
(a)

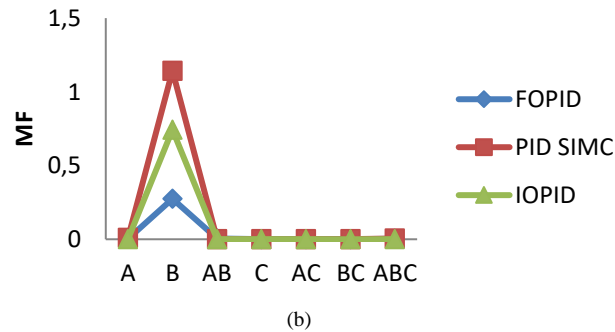
(b)

Figure 4. MF for robustness analysis of the control action: b) mean value and c) standard deviation.

V. CONCLUSIONS

This paper presented the design of a fractional order PID controller for a nonlinear system, and its statistical robustness analysis in the presence of external disturbances, parametric uncertainness, and random noise employing a factorial experimental design $2^3$. The FOPID controller was compared with an IOPID and SIMC PID controllers used to control a non-linear second-order system linearized through the input-output linearization technique. The factors used to measure robustness were uncertainty in the gains of the process, the presence of external perturbation and random noise in the feedback loop. The robustness analysis was performed by observing the effect of the factors on the input step response of the system and the control action of the feedback system. The results of the statistical robustness analysis showed that in the face of uncertainties in the process and the presence of external perturbations, the FOPID controller exhibits the best performance in the system step response and the control action. The robustness analysis shown that the FOPID controller is a good option for the control of process with parametric uncertainty in the presence of external perturbations, because reduce control effort required of the plant actuators, which translates into lower energy consumption for the system. Finally, the experimental design methodology used in this paper for the robustness analysis of control systems makes it possible to quantify and provide numerical evidence of the real performance of controllers in the presence of different factors that affect the closed loop behavior of the system.